\documentclass{Interspeech2024}

\title{%Critical Perspectives on 
Data Ethics and Practices of Human-Nonhuman Sound Technologies and Ecologies}

\interspeechcameraready 

\name[affiliation={1}]{Petra}{Jääskeläinen}
\name[affiliation={1}]{Elin}{Kanhov}
\address{
  KTH Royal Institute of Technology, Sweden
\email{ppja@kth.se, ekanhov@kth.se}
}

\keywords{nonhuman data ethics, data ethics, human-nonhuman interaction, human-animal interaction, data extractivism, technological mediation}
\begin{document}
\maketitle
% discuss in later paper: indirect vs direct consequences. 
\begin{abstract}
    Human-nonhuman sound interaction and technologies aim to bridge the gap of inter-species communication. While they emerge from attempts to understand and communicate with nonhumans, they also raise questions on the ethics of nonhuman data use, for example regarding the unintended consequences such data extraction can have to nonhumans. In this paper, we discuss power relations and aspects of representation in nonhuman data practices, and their potential critical implications to nonhumans. Drawing from prior research on data ethics and posthumanities, we conceptualize two challenges of nonhuman data ethics for the design of Human-Nonhuman Interaction (HNI) and technologies in sound ecologies. We provide takeaways for how sensitivities toward nonhuman stakeholders can be considered in the design of HNI in the context of sound ecologies. 
    \textcolor{white}{\footnote{Both authors have contributed equally to this paper.}}    \\
\end{abstract}

\section{Introduction}
%State the problem/the claims that are investigated: 
While research on human-nonhuman interaction is developing in many domains, including animal communication \cite{beecher2021} and human-computer interaction \cite{mancini2011, hirskyj2022}, so far little focus has been placed on the ethical aspects of nonhuman data use and data practices in the context of such interactions in sound ecologies. These ethical aspects has been raised previously in nonhuman philosophy and ethics, for example in terms of power structures between humans and nonhumans \cite{singer_animal_1975}, nonhuman representation \cite{Cesaresco2014-lk}, and labour \cite{barron}. All of these concerns can be directly projected to examine nonhuman data ethics and practices. 

%State what we mean by ethical aspects: Power relations, extractivism.
By \textit{human-nonhuman interaction (HNI)}, we refer in this paper broadly not only to inter-species communication and design of technologies for such purposes, but also to the actuality of humans living, both passively and actively, in constant interaction and relationality -- or entanglement -- with nonhumans \cite{barad_meeting_2007}. In this paper, we explicitly focus on living nonhuman entities (e.g. animals, plants, ecosystems) rather than, for example, technological companions \cite{haraway2016}. Humans interact with nonhumans simply by entering their habitat and observing their ways of life, without attempts of communicating. In this way, we take a relational \cite{west2021} environmental posthumanist perspective \cite{Haraway1987, barad_meeting_2007, braidotti_posthuman_2019, asberg2021} on the kinship between humans and nonhumans \cite{haraway2016}, and focus on the role of sound in such relational ecologies \cite{truax_world_1978,schafer_soundscape_1993,droumeva_sound_2019}. Within this context of HNI, our specific focus is therefore to examine \textit{the ethics of data practices} in nonhuman sound ecologies. These questions arise, for example, when we enter environments in which nonhumans reside; introduce technology into them; design technologies for inter-species interaction; and generally when we collect and use nonhuman data. There is a distinction to be made between ethics of data use and ethics of entering nonhuman environments for data collection, and we discuss both of these in this paper under the term data practices. Not all processes require both of these, and it is likely that they raise different ethical issues in practice.

In this work, we draw from data ethics, posthumanities, and sound ecologies literature to inform the use of data in the context of human-nonhuman interactions, asking the question: \textit{how do sonic entanglements relate to questions of power dynamics between humans and nonhumans, and how may technological mediation affect such dynamics?} By drawing from existing literature \cite{DIgnazio2023-io}, we outline two ethical challenges of nonhuman data use and practices: 1. Examining and Challenging Human-Nonhuman Power Structures, and 2. Examining the Nonhuman Data Representation and Labour. Thus, this paper contributes with providing critical perspectives on data ethics and power relations of human-nonhuman interactions in sound ecologies. We discuss potential benefits and concerns in how research in this domain can configure power relations between humans and nonhumans through data practices \cite{mezzadra_multiple_2017,crawford_atlas_2021,DIgnazio2023-io}, and how technology plays an active role in configuring these relations through the mediation of sound between humans and nonhumans. Lastly, we urge for further critical reflection on nonhuman data ethics in HNI and sound ecologies.

We first begin by situating HNI into relational sound ecologies. Subsequently, we place practices of data extraction into the wider context of knowledge production through sound, to then conceptualize what nonhuman data ethics can implicate.

\section{Background}

\subsection{Relational Sound Ecologies} \label{entangledsound}

Sound is part of relational \cite{west2021, Haraway1987, latour} ecologies that involve both humans and nonhumans. In the era of the Anthropocene, these entangled and relational more-than-human ecologies are often discussed in terms of how they contribute to sustainability, such as biodiversity and maintaining healthy ecosystems \cite{west2021}. We consider ``sound ecologies" to be any more-than-human system in which sound plays a role in the relationality between entities. A more-than-human onto-epistemology in posthumanist research advocates relational thinking \cite{west2021} and a decentralization of humans \cite{Haraway1987, latour}, for example in relation to other living entities. These perspectives have been informed by both de-colonial research on indigenous environmental relations (e.g. place-based onto-epistemologies) \cite{Kimmerer2015-km}, and feminist science and technology studies (STS) \cite{Haraway1987, barad_meeting_2007}. 

However, in the Western modernist scientific paradigm \cite[e.g.]{latour_science_1987, west2021, haraway_situated_1988} there is a strong tendency to study ``measurable" and ``modellable" aspects, often with insufficient sensitivity to nonhuman subjectivities \cite{debaise_what_2012}. When we think of these human-nonhuman relations -- specifically in the context of sound-technology-mediation -- we need to consider questions of how human interactions with nonhumans, and technologies that mediate these interactions, shape the nonhumans' reality, rather than approaching them from an anthropocentric perspective. In an attempt to de-centralize these anthropocentric perspectives, we can begin to ``de-colonize" and reconfigure our relation to nonhumans -- an effort that has become increasingly explored in technology interactions in recent years in the form of more-than-human technology design \cite[e.g.]{wakkary_things_2021, Nicenboim2020-fn, crawford_atlas_2021}. 

Humans are deeply entangled with other species in sound ecologies, and this involves a constant configuration of power relations between various human and nonhuman entities. This becomes particularly evident in studies of noise pollution \cite{slabbekoorn_effects_2018,sordello_evidence_2020}, where human ways of life not only affects the physical environment of living nonhuman entities, but also silences their sonic expressions, capabilities and realities. However, it is not only through such destructive practices that humans are engaged in sound ecologies. Turning to indigenous cultures, it is clear that humans have long been sonically entangled with nonhumans. For example, the Kaluli people of Papua New Guinea have a deep sonic and musical connection with their environment \cite{feld_sound_2012}. Such ways of knowing have, not least in the Western world, been undermined by rationalization in the modernist scientific paradigm. 

Furthermore, due to the differences in our make compared to living nonhuman entities, in certain aspects we are also very concretely detangled from each other's sonic realities. For example, humans are incapable of hearing infrasounds of breaking icebergs, whales and elephants, and ultrasounds of bats, mice and corals, as frequencies of such sounds lie outside of the human range of hearing \cite{bakker2022, nagel2024like}. By using technological tools and mediation, however, humans can become able to hear these sonic realities of other living entities. %By using the term ``able", we actively through the use of language attempt to reframe humans as not the species par excellence, but as lacking rich capabilities and ways of sensing that various nonhumans have. 

\subsection{Knowledge Production and Data in Sound Ecologies} \label{soundknowledge}

Knowing the world through sound offers information and sensory input that widely differ from visual inquires, which are often dominating the ways of knowing for humans. Thus, sonic imagination in itself can help us think beyond the visually dominated human-centred world \cite{schafer_soundscape_1993,wrightson2000introduction,tkaczyk_sonic_2020,ogorman_sounding_2023}. These visual ways of knowing are central also to other primates, which thus are naturally advantaged from sharing the same senses as humans in this human-centred world. As such, sonic perspectives can be understood as part of an embodied, embedded and situated knowledge practice \cite{haraway_situated_1988,braidotti_posthuman_2019}, where an ``acoustemology" \cite{novak_acoustemology_2015}, or acoustic epistemology, affords sensitivities towards nonhuman subjectivities beyond normative (Western rational) ways of knowing. This notion of embodied knowledge has been generally acknowledged in the design of technology in past decades \cite{dourish, hook2020, hook2021}, changing the way how technology design is approached.

To access the world of sound beyond using our ears, which, as we have already established, are limited in terms of range and sensitivity to certain levels of sound, we can turn to technology to ``enhance" and ``decode" sound ecologies. In fact, the digital revolution has offered new tools and methods for accessing nonhuman sound ecologies that has provided understandings for how complex such ecologies are \cite{bakker2022}. This affords not only new incentives for environmental conservation but also possibilities for inter-species communication. For such practices to be possible, however, the data that is recorded, or \textit{extracted} from ecological sites, must be manipulated so as to be intelligible to humans. 

A critical question therefore arises regarding what such processes of technologically enhanced entanglements induce, if we examine the power relations and focus on the subjectivities of nonhumans. While technological developments and capabilities enable further exploration of the sonic world and provide insight and deeper understanding of nonhuman realities, they also have the potential to disturb and change the natural habitats and behaviours of the nonhumans studied \cite{slabbekoorn_effects_2018}. As such, there is a danger that technology becomes a tool for extractivist practices toward nonhumans, serving the anthropocentric worldview and enforcing the contemporary power configurations that place humans as the locus. It is essential, then, that ethical reflection is directed toward the potential critical impact on nonhumans when such technologies are designed and introduced in these more-than-human configurations.

Furthermore, it is important to note that different research fields have varying motivations and intentions for their sonic data collection. This can be due to cultural, geopolitical, and institutional differences, and their ethical guidelines and practices often vary. For example, while animal behavioral research has the intent to understand nonhumans, technology engineering research has a primary interest in advancing technological development, and artistic practice might work with nonhuman data in creative dialogue with society. In summary, the human-nonhuman sound interaction is a very diverse field of practices, and the data ethics practices of each specific case should be examined carefully.

\section{Conceptualizing Nonhuman Data Ethics} \label{}
Exploring these critical questions and impacts on nonhumans further, we turn to feminist data ethics literature \cite[e.g.]{DIgnazio2023-io} as a perspective to understand how power relations are constructed through data and data practices. Bringing this together with other literature that examines power relations between humans and nonhumans (such as speciesism \cite{singer_animal_1975} and human-animal media studies \cite{Cesaresco2014-lk}), we argue that data practices involving nonhumans are actively configuring inter-species power relations. In this section, we draw on this research to conceptualize important dimensions that need to be examined in terms of ethics of nonhuman data and sound technology practices. 

The principles of \textit{data feminism} are intended to re-think and reconfigure power relations in the context of human data practices. In regards to more-than-human sound ecologies, we can apply the same principles to examine power relations of nonhuman data use and practices -- a connection that feminist environmental posthumanities research has more widely built on to examine questions that relate to human-nonhuman relations \cite{asberg2021}. There are seven feminist principles for working with data, which we will examine in the context of nonhuman data in sound ecologies. These are; 1. Examine power, 2. Challenge power, 3. Elevate emotion and embodiment, 4. Rethink binaries and hierarchies, 5. Embrace pluralism, 6. Consider context, 7. Make labour visible  \cite{DIgnazio2023-io}. Examining power concerns the need to critically investigate the power configurations that relate to data and data practices, and challenging power means taking concrete steps of re-configuring the identified power imbalances. Elevating embodiment highlights the earlier discussed need to expand the knowledge-making to its embodied situatedness. Rethinking binaries and hierarchies can help change the way information is conceptualized, leading into embracing pluralism which encourages diverse ways of knowing, communicating and being. Consideration of context refers to acknowledging the situated context of each case, and lastly, making labour visible concerns tracing and exposing all the labour that takes place in data practices. We now project these principles onto the case of HNI in sound technologies and ecologies. 

\subsection{Examining and Challenging Human-Nonhuman Power Structures}
Feminist data ethics advocate for firstly examining prevailing power structures, to then actively challenge them. Transferring this principle onto the design of HNI sound technologies, researchers should consider how power is configured between various human and nonhuman stakeholders with these technologies, and how sound technologies can be re-imagined in ways that the nonhuman stakeholders gain more power and agency. These aspects urge the designers and developers of the technologies to think about on whose terms the technology is designed and who is benefiting from it in the long term. Relevant questions to ask in this context are: \textit{how is power configured between various human and nonhuman stakeholders in the technological configurations}, and \textit{how can these technologies be radically re-imagined in a way that the nonhuman stakeholders gain more power}? These questions probe designers and developers of the technologies to think about on whose terms the technology is designed and who is benefiting from it in the long term. In a practical sense, approaches such as mapping the critical and positive stakeholder (nonhuman) concerns can be incorporated in processes of reflecting on such questions. These types of methods have recently started to emerge in HCI research \cite[e.g.]{nunes-poikolainen2024}. Thus, there surfaces a need to explore more methods that can be used in technology and data practices for developing sensitivities to nonhuman stakeholders.

Considering on whose terms the technology is designed, it is important to study data practices on a larger scale. This concerns, for example, what kind of practices and types of data are dominating the landscape in HNI. One of the central aspects that characterizes human data practices and, more widely, practices of designing technology, is the aspire to \textit{decode}, \textit{systematize}, and \textit{model} \cite{latour_science_1987, haraway_situated_1988}. Designers and developers should consider how these processes of technological mediation affect the type of information that is mediated, and what is gained or lost when we try to organize nonhuman sounds in ``human ways". Prior studies have explored data surveillance and data extraction in the context of various nonhumans, for example discussing how modeling and rationalizing can lead to harmful outcomes for the nonhumans \cite{tironi}. Also, studies demonstrate how data practices configure new environments and nonhuman-environmental relations, and how such practices give voice to various ``monitored" nonhumans (e.g. animals, plants) \cite{gabrys2016program,farley_2018}. As humans attempt to monitor, record, decode, analyse and even communicate with nonhumans, we need to ask on whose terms these (inter-)actions are practiced.

It is also crucial to examine processes of intervention, and how human and technological presence in nonhuman habitats may affect the nonhuman ecologies, related to the third principle of elevating emotion and embodiment. As discussed, sound and particularly vocalization plays a role in the power dynamic between humans and nonhumans. For instance, cats vocalize in a particular way when engaging with humans, and animals that are taken out of their natural habitats can start vocalizing more intensely as an sign of dependence on human caretakers. By practicing empathy toward the nonhuman and fully engaging in sensitive and embodied listening, designers and developers can ``make kin", e.g. reflect and reconfigure our relation to nonhumans \cite{haraway2016}. Furthermore, we need to fully understand the long-term implications of placing technological artefacts (mics, sensors, transmitters, etc.) in nonhuman sound ecologies, and how the nonhumans change and adapt to these. In the posthumanist literature, it has been explored how the human has co-evolved with technology through the concept of the cyborg \cite{haraway_simians_1991}. Like humans, nonhuman entities are not immune to technological influence, and it can be argued that they are also in a cyborg relationship with their (technological) environments \cite{gabrys2016program}. Yet, they have less power in giving consent to being so. From a sound ecology perspective, data collection practices can also involve introducing sounds to wild environments, which calls for ethical reflection on the impact of our data practices on sound ecologies. For instance, researchers may purposefully introduce sounds to lure birds or other species into communication, or simply produce sounds by talking, walking, and using vehicles.

Following de-colonial science and technology practices \cite{Costanza-Chock2020-jb, garcia2021}, researchers can further ask whether we always have the right to enter a nonhuman habitat for the sake of scientific and technological advancement. This question urges us to examine our human privileges, and our role as ``nonhuman colonizers" that use technology as a tool for colonization. While these issues have surfaced often in de-colonial data studies \cite{mejias_data_2024}, they have not been examined in depth when it comes to HNI. Thus, we urge these questions to become an integral part of data ethics in HNI sound ecologies.

\subsection{Examining Nonhuman Data Representation and Labour}
Turning to the principles of rethinking binaries and hierarchies, as well as embracing pluralism, another relevant dimension of data practices relates to representation, which is commonly discussed in human data ethics \cite{DIgnazio2023-io}. This concerns what and who is represented in data collection and analysis, which in human terms is discussed in aspects of gender and race, for example. Transferring this notion to the context of nonhuman data practices, we can ask: \textit{which species are studied and which are not}, and \textit{what data is dominating in the data practices}? This also raises questions on what the critical implications are for various species when they are represented in differing ways. For example, a lack or excess of representation of certain species may affect their everyday life and experiences, as some species might be considered more ``worth" studying than others (e.g. \cite{nagel2024like}). Furthermore, we can also ask what implications there are if the species are represented and discussed in a certain normative way. As an example, when animals are represented in human culture (media), people might be more likely to approach and interact with certain familiar species in the wild or sympathize more with such species which can have direct consequences for their livelihoods and environments. Similarly, nonhumans that are deemed ``hostile" can be treated in very different and non-caring ways by humans -- or even completely disregarded and excluded from conservation. 

Diversifying representations not only applies to the data itself, but also to multiple ways of knowing and making knowledge (as discussed in Section \ref{soundknowledge}). This can be done by challenging the predominant ways of doing research in HNI and seeking to diversify such practices. These remarks urge the designers and developers of HNI technologies to reflect carefully on data practices, collection, and use in terms of how the data is manipulated; what forms it takes; what ways of knowing it promotes; and ultimately, what ways of knowing are prioritized and dominating the data practices. Furthermore, such diversifying can be cultivated by attuning to ways of being and knowing that are currently overlooked or underrepresented. These aspects urge us to fine-tune into and examine more carefully the contexts in which the data exists, is produced, and understood.

Related to considerations of context and making labour visible, we wish to emphasize the need to acknowledge nonhuman labour in collection of data and design of the technology. Most often in HNI, nonhumans are contributing their data without having a choice to do so. It is therefore also relevant to consider whether they should be compensated for that data extraction, and whether there are ways of asking nonhumans for consent of use. In animal ethics \cite{singer_animal_1975} and environmental ethics \cite{goodpaster1978} it has been argued that ethical consideration should be attributed to nonhumans following their unique needs. For example, species with similar needs call for similar consideration and care, as a principle of equal treatment. When this is applied to labour and data ethics, we can consider different nonhuman species to do differing types of labour -- actions or behaviors -- in producing data and interacting with humans and technology. Furthermore, we can anticipate a need for them to be compensated differently from this labour, following each species' unique needs and interests. This raises challenging questions about how such compensation should take place. For example, if we compensate zebra finches species members with plant seeds, it can be seen as their species-specific interest. At the same time, we might contribute to domestication of the species and further inter-species colonization. Furthermore, we might overlook the individual preferences and variability of specific species members \cite{aaltola2006animal}.

Lastly, the labour that both humans and nonhumans engage in is actively shaping the earlier discussed representations of nonhumans by rendering some species more visible than others. Reflecting on how such nonhuman data labour can be practiced on ethical terms is therefore of critical importance -- in a similar way to how the handling of human data is becoming an increasing concern in all parts of digital society \cite{mejias_data_2024, crawford_atlas_2021, DIgnazio2023-io}.  

\section{Conclusion}

In this paper, we have discussed critical questions in regard to the data ethics of human-nonhuman in sound technologies and ecologies. Drawing from feminist and de-colonial data ethics, posthumanities, and sound ecologies literature, we have conceptualized sound ecologies as relational sites in which knowledge production and data practices coincide. We have provided two concrete areas to examine when it comes to nonhuman data ethics (1. Examining and Challenging Human-Nonhuman Power Structures, and 2. Examining Nonhuman Data Representation and Labour). We discussed related challenges through concrete examples, and reflected on what unintended consequences such data practices can have to nonhumans. We aim for this paper to spark discussion on data and sound technology practices in the communities that design human-nonhuman interactions, and urge for the VIHAR community to examine these data ethics questions in further depth in the future.

\section{Acknowledgements}
This work was partially supported by the Wallenberg AI, Autonomous Systems and Software Program – Humanities and Society (WASP- HS) funded by the Marianne and Marcus Wallenberg Foundation, and a project that has received funding from the European Research Council (ERC) under the European Union’s Horizon 2020 research and innovation programme (Grant agreement No. 864189).

\bibliographystyle{IEEEtran}
\bibliography{mybib}

\end{document}